\documentclass[iop]{emulateapj}
\slugcomment{Accepted to the Astrophysical Journal}
\usepackage{amsmath}
\usepackage{color}

\shorttitle{Atmospheric circulation of Large Program planets}
\shortauthors{Kataria et al.}

\begin{document}

\title{The atmospheric circulation of a nine-hot Jupiter sample: Probing circulation and chemistry over a wide phase space}

\author{
Tiffany Kataria\altaffilmark{1}, David K. Sing\altaffilmark{1}, Nikole K. Lewis\altaffilmark{2}, Channon Visscher\altaffilmark{3},
Adam P. Showman\altaffilmark{4}, Jonathan J. Fortney\altaffilmark{5}, Mark S. Marley\altaffilmark{6}  }

\altaffiltext{1}{Astrophysics Group, School of Physics, University of Exeter, Stocker Road, Exeter EX4 4QL, UK; \email{\bf tkataria@astro.ex.ac.uk}}
\altaffiltext{2}{Space Telescope Science Institute, 3700 San Martin Drive, Baltimore, MD 21218, USA}
\altaffiltext{3}{Department of Chemistry, Dordt College, Sioux Center, Iowa 51250, USA}
\altaffiltext{4}{Department of Planetary Sciences and Lunar and Planetary Laboratory, The University of Arizona, Tucson, AZ 85721, USA}
\altaffiltext{5}{Department of Astronomy \& Astrophysics, University of California, Santa Cruz, CA 95064, USA}
\altaffiltext{6}{NASA Ames Research Center 245-3, Moffett Field, CA 94035, USA}

\begin{abstract}
We present results from an atmospheric circulation study of nine hot Jupiters that comprise a large 
transmission spectral survey using the Hubble and Spitzer Space Telescopes.  These observations 
exhibit a range of spectral behavior over optical and infrared wavelengths which suggest diverse 
cloud and haze properties in their atmospheres.  By utilizing the specific system parameters for 
each planet, we naturally probe a wide phase space in planet radius, gravity, orbital period, 
and equilibrium temperature.  First, we show that our model ``grid" recovers trends shown in traditional 
parametric studies of hot Jupiters, particularly equatorial superrotation and increased day-night temperature contrast with increasing 
equilibrium temperature. We show how spatial temperature variations, particularly
between the dayside and nightside and west and east terminators, can vary by hundreds of K, which
could imply large variations in Na, K, CO and $\rm{CH_4}$ abundances in those regions.  These chemical variations
can be large enough to be observed in transmission with high-resolution spectrographs, such as ESPRESSO on
VLT, METIS on the E-ELT, or with MIRI and NIRSpec aboard JWST.  We also compare theoretical 
emission spectra generated from our models to available Spitzer eclipse depths for each planet, 
and find that the outputs from our solar-metallicity, cloud-free models generally 
provide a good match to many of the datasets, even without additional model tuning.    
Although these models are cloud-free, we can use their results to understand
 the chemistry and dynamics that drive cloud formation in their 
atmospheres.

\end{abstract}


\section{Introduction}

After nearly two decades of milestone discoveries in exoplanet science, current and future observational efforts 
seek to discover and characterize smaller and cooler planets at larger orbital distances.  Even so, the close-in, 
tidally-locked ``hot Jupiters" remain an important population for exoplanet characterization--they are still the best
 targets for probing atmospheric properties, refining observational techniques, and expanding current theory, all 
 of which can be extended to smaller planets.  
 
  \begin{deluxetable*}{lccccccccccc}
%\begin{deluxetable}{lll}
\tabletypesize{\footnotesize}
\tablecaption{Planetary and stellar parameters of the HST Large Program (LP) planets modeled in our study.  
See main text for definitions of columns 8-12.}
\tablewidth{0pt}
\tablehead{
\colhead{Planet} & \colhead{$\rm{R_p}$} & \colhead{${T_{orb}=T_{rot}}$} & \colhead{${ \Omega}$} & \colhead{${ g}$} & 
\colhead{${ T_{eq}}$} & \colhead{$F_\star$} & \colhead{$H$} & \colhead{${ R_o}$} & \colhead{${ N_{jets}}$} & 
\colhead{${ L_D}$} & \colhead{${ L_{\beta}}$} \\
\colhead{} & \colhead{(${\rm R_{J}}$)} & \colhead{(Earth days)} & \colhead{(${\rm 10^{-5}~s^{-1}}$)} & 
\colhead{(${\rm m~s^{-2}}$)} & \colhead{(K)} & \colhead{(${\rm W~m^{-2}}$)} & \colhead{(km)} & \colhead{} & 
\colhead{} & \colhead{${\rm(\times10^7~m)}$} & \colhead{($ {\rm km}$)} }
%planet &      Rp &      Trot &  Omega & g & Teq & Fstar & H &    Ro &    Njets & Ld
\startdata
HAT-P-12b & 0.96 & 3.21 & 2.263 & 5.62  & 963 & 1.917e5 & 648 & 0.82 & 1.28 & 5.4 & 16.7\\
WASP-39b & 1.27 & 4.06 & 1.793 & 4.07  & 1117 & 3.548e5 &1037 & 0.92 & 1.35 & 6.5  & 20.3\\
WASP-6b & 1.22 & 3.36 & 2.164 & 8.71  & 1145 & 3.452e5 & 497 & 0.89 & 1.26 & 7.2 & 19.6\\
HD 189733b & 1.14 & 2.22 & 3.279 & 21.40  & 1201 & 4.632e5 &  212 & 0.67 & 1.59 & 5.1 & 16.4 \\
HAT-P-1b & 1.32 & 4.46 & 1.631 & 7.46  & 1322 & 7.016e5 & 670 & 1.32 & 1.17 & 8.0 & 24.8\\
HD 209458b & 1.36 & 3.52 & 2.066 & 9.35  & 1448 & 1.001e6 & 585 & 1.26 & 1.31 & 7.4 & 24.6 \\ 
WASP-31b & 1.55 & 3.40 & 2.139 & 4.56  & 1575 & 1.399e6 & 1305 & 1.06 & 1.40 & 7.9 & 24.0\\
WASP-17b & 1.89 &  3.73 & 1.950 & 3.57  & 1738 & 2.140e6 & 1840 & 1.09 & 1.43 & 9.4  & 26.9 \\
WASP-19b & 1.41 & 0.79 & 9.217 & 14.21  & 2050 & 4.392e6 & 545 & 0.32 & 2.59 & 3.9 & 12.5
\enddata
\tablecomments{The values for $R_p$, $T_{orb}$, $T_{rot}$, $\Omega$, $g$, $T_{eq}$,
 $F_\star$ and $H$ were adopted from values in \cite{sing+2016} and references therein.  The values for $R_o$, $N_{jets}$, 
 $L_D$, and $L_{\beta}$ were derived from GCM outputs.  See text for more details.} 
\label{LP_params}
\end{deluxetable*}

Because many hot Jupiters transit their host stars, observations during transit (when the planet passes in front 
of its star) through secondary eclipse (when the planet passes behind its star) have allowed us to characterize their 
atmospheres.  Already, dozens of hot Jupiters have been observed with the Spitzer Space Telescope, Hubble 
Space Telescope (HST), and many ground-based facilities \citep[e.g.,][]{charb+2002,deming+2005,redfield+2008}.  Our insights will only deepen 
with the advent  of eclipse mapping, which can place constraints on both the longitudinal and latitudinal temperature structure
  \citep[e.g.,][]{dewit_etal_2012,nikolov+sainsbury_2015}.  Despite occupying a 
small region of exoplanet parameter space, these observations have shown that hot Jupiters exhibit a wide diversity in spectral 
 properties and atmospheric compositions. 

Recently, \cite{sing+2016} presented an ensemble of transmission spectra from a large HST 
observational program of ten hot Jupiters (HD 189733b, HD 209458b, WASP-6b, WASP-12b, WASP-17b, WASP-19b, WASP-31b, WASP-39b, 
HAT-P-1b, and HAT-P-12b) using STIS and WFC3, as well as the Spitzer 3.6 and 4.5 $\mu$m IRAC channels (HST GO-12473; P.I. Sing).  Their spectra show a range of water and alkali abundances, as well as Rayleigh scattering slopes at near-UV and optical wavelengths, 
that indicate a diversity in cloud and haze properties \citep[see also][]{pont+2013,sing+2013,huitson+2013,wakeford+2013,
nikolov+2014,nikolov+2015,sing+2015}.  More importantly, the water abundance for each planet correlates with atmospheric type, suggesting that hot Jupiters are not depleted in primordial water during formation, but simply that clouds and hazes obscure its spectral feature. 

General circulation models (GCMs) are useful tools for understanding the nature of these clouds and hazes, as the
 three-dimensional wind and temperature structure sets their formation and transport.  
Here we present cloud-free GCMs coupled to a nongray radiative transfer scheme
 for each Large Program (LP) planet.  Unlike other parametric circulation studies of hot Jupiters, which typically model one or two 
individual planets, this study is the first of its kind to compare circulation models for nine individual planets.  Such 
a large ensemble of hot Jupiters naturally allows for a comparative study over a wide parameter space in 
planet radius, gravity, orbital period, and equilibrium temperature. 

We should stress that the primary focus of this paper is not to provide a detailed analysis of the transmission 
spectra themselves.  Rather, this paper serves as the first step to interpreting the ensemble of transmission spectra 
by exploring each planet's three-dimensional atmospheric temperature structure and dynamics, which set the formation of the
clouds and hazes that are inferred from the data.  We do, however, compare our models
to observed dayside thermal emission observations, as those datasets are directly related to each planet's 
thermal structure.  Furthermore, we stress that the main objective of this study is not to analyze
each individual planet on its own; rather, we aim to use this ``grid" of planets to probe trends in dynamics,
temperature, and chemistry across the entire sample.    In Section \ref{section:modelsetup} we describe our model setup, the SPARC/MITgcm.
In Section \ref{section:results} we present the major results of our comparative study.  Section \ref{section:discussion} provides a discussion of 
these results, and we conclude in Section \ref{section:conclusions}.
 
\begin{figure*}
   \centering
   \includegraphics[trim = 0.2in 0.0in 0.15in 0.0in, clip, width=0.95\textwidth]{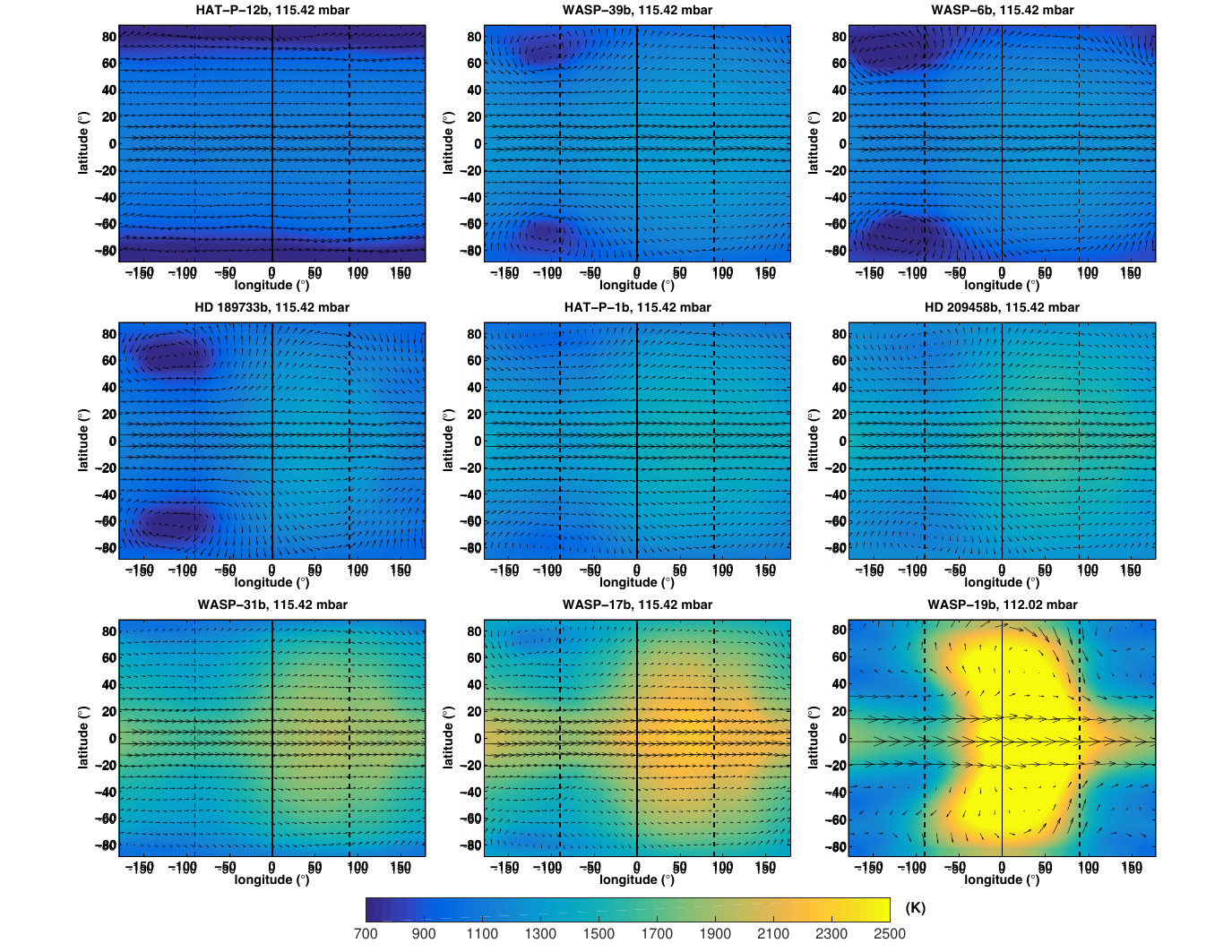}\\
   \caption{Wind/temperature profiles for each LP target at a pressure of 100 mbar.  Panels are organized by increasing planetary equilibrium 
   temperature, $T_{eq}$, and are plotted on the same colorscale.  Each vertical solid line denotes the substellar longitude, while dashed vertical lines 
   denote the western and eastern terminators, respectively. }\label{uvt_plots}
\end{figure*}
%%%%%%%%%%%%%%%%%%%%%%%%%%%%%%%%%%%%%%%%%%%%%%%%%%%%%%%%%%%%%%

\section{Model Setup}\label{section:modelsetup}
To model each planet's atmospheric circulation, we utilize the Substellar and Planetary 
Radiation and Circulation (SPARC) model \citep{showman+2009}.  The SPARC model couples the general 
circulation model maintained at the Massachusetts Institute of Technology \citep[the MITgcm,][]{adcroft+2004} 
with a plane-parallel, two-stream version of the multi-stream radiation code developed by \cite{marley+mckay1999}. 
The MITgcm is a finite-volume code that solves the three-dimensional primitive equations on a staggered Arakawa C 
grid \citep{arakawa+lamb1977}.  We use a curvilinear coordinate system called the ``cubed-sphere"; unlike a 
traditional latitude-longitude grid, the cubed-sphere lacks singularities at the poles, allowing for longer timesteps 
throughout the domain.  We maintain numerical stability using a fourth-order Shapiro filter on the time derivatives 
of the wind and temperature at each timestep; this smooths the grid-scale variations while minimally affecting 
the flow at large scales. 

The radiative transfer (RT) code solves the two-stream radiative transfer equations and employs the correlated-$k$ 
method \citep{goody+1989,marley+mckay1999} over 11 spectral bins \citep{kataria+2013}, which retains most of the accuracy of full 
line-by-line calculations while drastically increasing computational efficiency.  Opacities are calculated assuming 
local thermodynamic and chemical equilibrium for each pressure-temperature ($p$-T) point, using the solar photospheric elemental 
abundances of \cite{lodders2003}.  The coupling of the dynamical core and radiative transfer scheme allow for 
the self-consistent calculation of the heating and cooling rates of the atmosphere.  At each grid point, the MITgcm 
calculates the wind and temperature fields, which are used by the RT scheme to calculate the upward and downward 
fluxes at each pressure layer.  These fluxes are used to update the heating and cooling rates, which are then used 
by the MITgcm.  

The SPARC model is already fully operational and has been extensively used for hot Jupiters \citep{showman+2009,
kataria+2013, parmentier+2013, showman+2013,lewis+2014, kataria+2015,showman+2015}, sub-Neptunes \citep{lewis+2010}, and 
super Earths \citep{kataria+2014}.  We derive our input spectrum for each parent star from the PHOENIX stellar atmosphere models \citep{hauschildt+1999}.

We focus our modelling on nine of the ten individual planets that comprise the LP 
sample.  Their system parameters are listed in Table~\ref{LP_params}.  We leave a detailed study of the circulation of 
WASP-12b as future work, noting that with its large equilibrium temperature ($\sim$2500 K), it is far out of the parameter 
space probed in this study. Table~\ref{LP_params} also includes values that help diagnose the dynamics (columns 8-12).  
We briefly discuss them here; for a more complete discussion of the parameters and their interpretation, 
see \cite{showman+2010}.  

First, we calculate each planet's pressure scale height, $H$, which is generally defined as 
\begin{equation}
H = \frac{kT}{mg}
\end{equation}
where $k$ is the Stefan-Boltzmann constant, $T$ is the temperature, $m$ is the mean-molecular weight 
and $g$ is the planetary gravity.  Here, we calculate a reference scale height using the equilibrium temperature, $T_{eq}$.
The Rossby number, $R_o$, is a dimensionless parameter which diagnoses the 
importance of rotation on the dynamics.  It is defined as 

\begin{equation}
R_o = \frac{U}{fL}
\end{equation}
where $U$ is a characteristic wind speed, $f$ is the Coriolis parameter, $2\Omega\sin\phi$, where $\phi$ is the latitude, 
and $L$ is a characteristic length scale (in this case, the scale height, $H$).  We assume the characteristic wind speed 
to be the root-mean squared (RMS) velocity\footnote{the RMS velocity was calculated as prescribed in \cite{lewis+2010},  $V_{rms}(p) = \sqrt{\frac{\int (u^2+v^2)dA}{A}}$, where $u$ is the zonal velocity and $v$ is the meridional velocity.} at 100 mbar from our simulations, approximately the level of the infrared (IR) photosphere.  
We also calculate two dynamical length scales for each planet: the equatorial deformation radius, ${L_D}$, which 
describes the characteristic length scale over which the atmosphere geostrophically adjusts to large-scale phenomena, and the Rhines scale, $L_\beta$, which defines the length scale over which a flow transitions to zonally-banded flow 
(that is, the length scale over which the flow reorganizes into jets)\citep{holton_1992}.  At the equator, these are defined as

\begin{equation}
L_D = \Big(\frac{NH}{\beta}\Big)^{1/2}
\end{equation}
and 
\begin{equation}
L_\beta= \pi\Big(\frac{U}{\beta}\Big)^{1/2}
\end{equation}
where $N$ is the Brunt-V\"{a}is\"{a}l\"{a} frequency, and $\beta={df}/{dy}$ is the beta parameter, the variation in the 
Coriolis force with latitude.  At the equator, $\beta = 2\Omega\cos\phi/\rm{R_p}$, where $\rm{R_p}$ is the planet radius.  

 \cite{showman+polvani_2011} have shown that the equatorial jet width should be set by $L_D$. 
We can then use our formulation of $L_D$ to derive an expression for the number of jets expected in each atmosphere by dividing the planet radius
by the deformation radius,

\begin{equation}
N_{jets} \sim \Big(\frac{2\Omega \rm{R_p}}{NH}\Big)^{1/2}
\end{equation} 
We also include each planet's  
orbital period, $T_{orb}$, and rotation period, $T_{rot}$, which are assumed to be equal (i.e., synchronous rotation).  

For all planet models, we assume an atmospheric metallicity of 1$\times$ solar. Models 
for planets cooler than WASP-19b have a horizontal resolution of C32 (128$\times$64 in longitude and latitude, respectively)
and a vertical resolution of 53 pressure levels, evenly spaced in log pressure, that extend from a mean pressure of 200 bars 
at the bottom to 2 $\mu$bar at the top.  We run the circulation model of WASP-19b at lower resolution to maintain numerical stability; 
it has a horizontal resolution of C16 (63$\times$32) and a vertical resolution of 46 pressure levels, evenly spaced in log 
pressure from a mean pressure of 1000 bars at the bottom to 0.2 mbar at the top.  We have shown in a previous study 
\citep{kataria+2015} that low resolution models still capture the bulk dynamical structure of hot Jupiters; therefore, the model 
of WASP-19b is sufficient for our comparison study.

\begin{figure*}
   \centering
   \includegraphics[trim = 0.3in 0.1in 0.25in 0.0in, clip, width=0.95\textwidth]{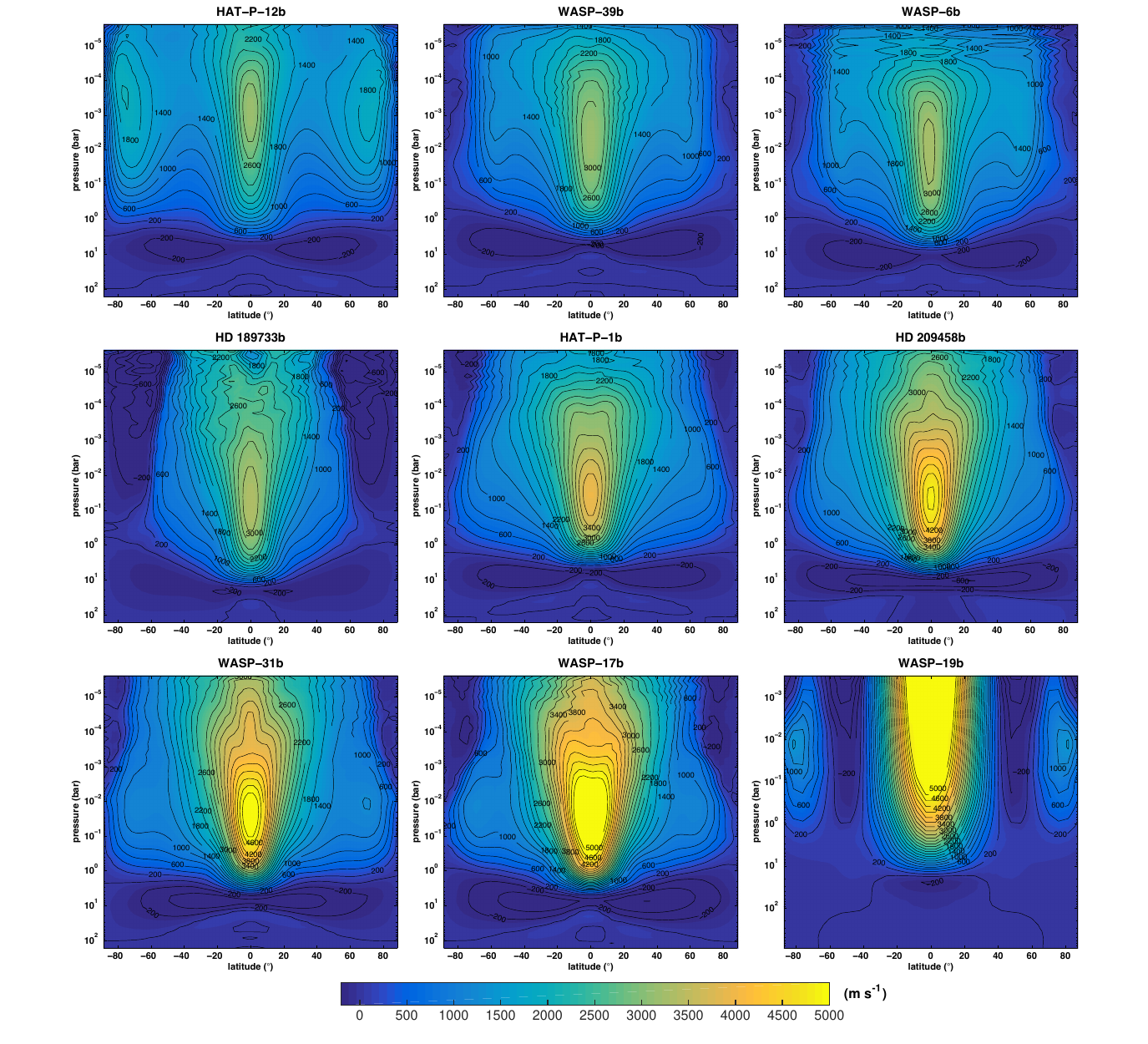}\\
   \caption{Zonal-mean zonal wind for each individual LP target .  Panels are organized by increasing planetary equilibrium temperature, $T_{eq}$, 
   and are plotted on the same colorscale.}\label{uz_plots}
\end{figure*}

%%%%%%%%%%%%%%%%%%%%%%%%%%%%%%%%%%%%%%%%%%%%%%%%%%%%%%%%%%%%%%

\section{Results}\label{section:results}
\subsection{Circulation regime}\label{subsection:circregime}

Previous parametric studies \citep[e.g.,][]{perezbecker+showman2013,showman+2015} have shown that with increasing stellar irradiation, 
the atmospheric circulation of hot Jupiters transitions from a regime that is dominated by banded zonal flow with small 
day-night temperature contrasts, to a regime with strong day-night flow and large day-night contrasts\footnote{This trend 
exists even in the absence of friction, which would further inhibit day-night flow and also lead to large day-night contrasts.  
The circulation models presented here do not include drag, and we therefore do not discuss their effects in further detail.}.  
We recover this trend in our ``grid" of models for each individual planet when we sort their wind/temperature profiles and 
zonal-mean zonal wind profiles by increasing equilibrium temperature, ${T_{eq}}$ (Figures \ref{uvt_plots} and \ref{uz_plots}).   
When comparing the coolest hot Jupiter in our sample, HAT-P-12b, with the hottest planet in our sample, WASP-19b, at a given pressure 
of 100 mbar (within the range of observable pressures), HAT-P-12b exhibits little to no day-night temperature variation 
($<$50 K) with weak zonal flow, while WASP-19b has a very large temperature contrast ($\sim$800-1000 K, Figure \ref{uvt_plots}) and strong
equatorial flow. 

Despite these differences in day-night temperature structure, the prevailing wind direction at photospheric pressures is 
eastward for all planets in our sample, a consequence of each planet's equatorial superrotation (Figure \ref{uz_plots}). 
This superrotation arises from the large day-night forcing each planet receives throughout its orbit, which excite Rossby 
and Kelvin waves and induce eddy phase tilts that interact with the mean flow to transport eastward angular momentum 
to the equator \citep{showman+polvani_2011}.  This produces the chevron-shaped hotspot seen in wind and temperature 
profiles of HD 189733b, HAT-P-1b, HD 209458b, WASP-31b, WASP-17b, and WASP-19b (Figure \ref{uvt_plots}).  
The peak speeds of the equatorial jets increase with increasing  ${T_{eq}}$, from ${\rm\sim3~km~s^{-1}}$ for 
HAT-P-12b to ${\rm\sim6~km~s^{-1}}$ for WASP-19b.  Furthermore, each wind/temperature map shows that the 
western terminator is cooler than the eastern terminator (left and right dashed lines, respectively), which could imply 
different chemical properties across each limb (see Section \ref{section:chemistry}). 

Interestingly, HAT-P-12b, HD 189733b and WASP-19b each have multiple jets in their atmospheres.  In addition to an 
equatorial jet, HAT-P-12b has two eastward jets at high latitudes ($\pm80^{\circ}$) with speeds exceeding 1 ${\rm km~s^{-1}}$, 
while HD 189733b has two westward jets at mid-latitudes ($\pm60^{\circ}$) with speeds of 600 ${\rm m~s^{-1}}$.  
WASP-19b has both high-latitude ($\pm80^{\circ}$) eastward jets exceeding 1 ${\rm km~s^{-1}}$ and mid-latitude 
($\pm50^{\circ}$) westward jets exceeding 200 ${\rm m~s^{-1}}$. This is a result of each planet's comparatively 
fast rotation rate and small Rossby deformation radius (Table \ref{LP_params}), which allow for the formation of multiple jets. 
 The effect of rotation rate and orbital distance (and similarly, stellar flux, $F_\star$) on atmospheric circulation has been 
 explored in a number of parametric studies of hot Jupiters  \citep{kataria+2013, showman+2015}. In particular, \cite{showman+2015} 
 explore planets over a wide range of stellar fluxes and rotation rates, from ``cold" to ``hot" ($1.16\times10^4$ to $4.65\times10^5~{\rm W~m^{-2}}$) and from ``slow" to ``fast" ($\rm{8.264\times10^{-6}~to~1.322\times10^{-4}~s^{-1}}$). While this study does not probe as large a parameter space, we can place our model results for these three planets in the context of \cite{showman+2015}. HD 189733b is identical to the 
 ``H$\Omega_{\rm med}$" case, while HAT-P-12b has a similar 
 rotation rate to HD 189733b, but a smaller stellar insolation (Table \ref{LP_params}). Therefore, HAT-P-12b falls in the phase space between the 
 ``hot" and  ``warm"  moderately fast rotating hot Jupiters, where the flow transitions from superrotation at the equator to the high 
 latitudes \citep[see Figs. 3 and 4 in][second/third rows, middle panels]{showman+2015}.  WASP-19b is both highly 
 irradiated and a fast rotator, and therefore its atmosphere forms multiple high latitude jets.  

%%%%%%%%%%%%%%%%%%%%%%%%%%%%%%%%%%%%%%%%%%%%%%%%%%%%%%%%%%%%%%
\subsection{Three-dimensional temperature considerations}

\begin{figure*}
   \centering
   \includegraphics[trim = 0.3in 0.1in 0.3in 0.0in, clip, width=0.95\textwidth]{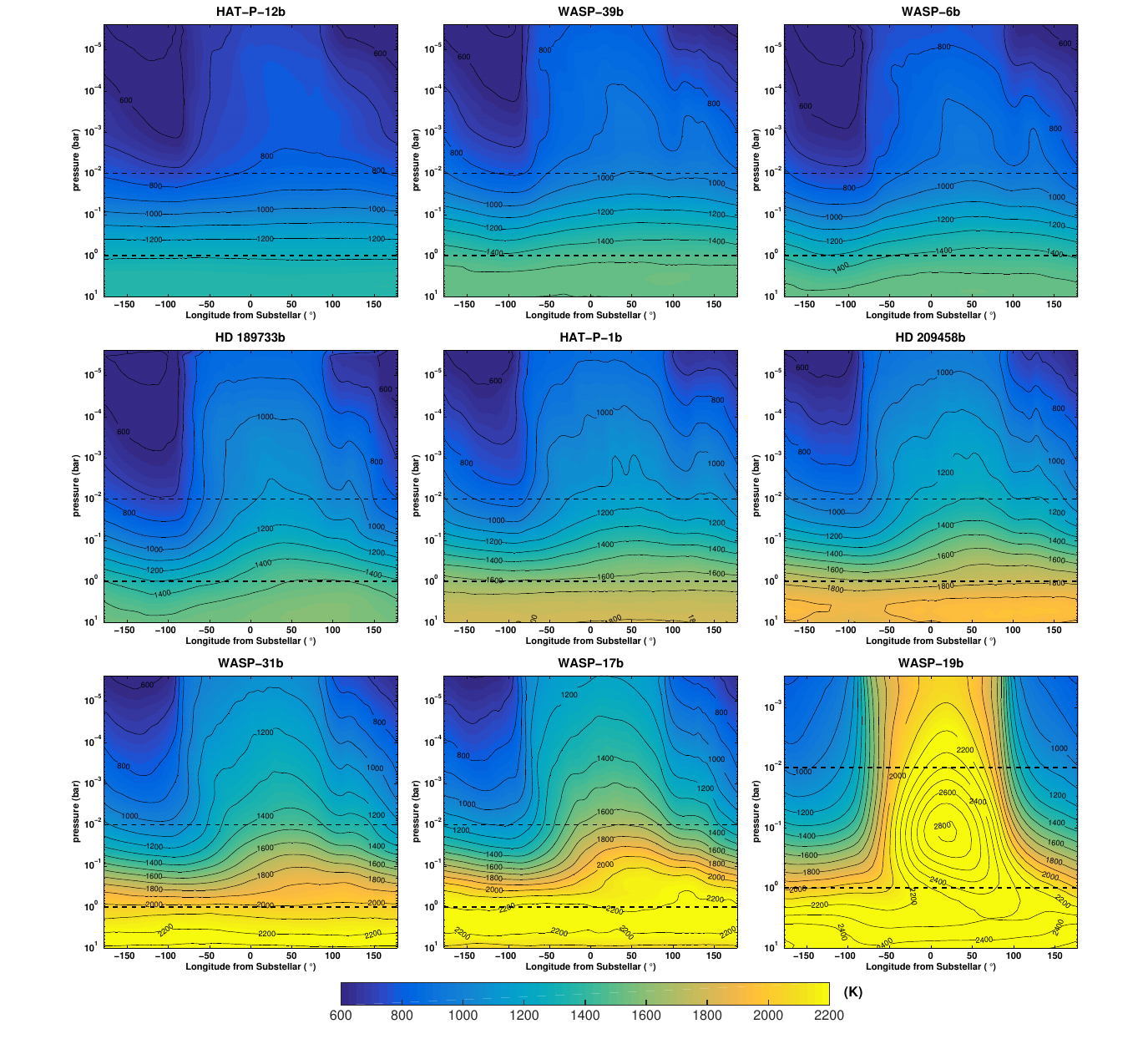}\\
   \caption{Latitudinally-averaged temperature as a function of pressure and longitude from the substellar longitude for each individual LP planet.  Panels are organized by increasing planetary equilibrium temperature, ${T_{eq}}$, and are plotted on the same colorscale. These average temperatures are weighted by cos $\phi$, where $\phi$ is latitude; this is equivalent to weighting each grid point by its projection angle toward an observer at the equator.  Black dashed lines highlight the pressures typically probed in emission.}\label{temp_lon_maps}
\end{figure*}

While transiting planets offer many observational strategies by which we can characterize their atmospheres, one should 
be careful to consider that each technique probes different spatial regions of the planet, from the planet's terminator or limb 
during transit, to the planet's dayside during secondary eclipse, to time-varying longitudes for full phase curve observations.  
The temperatures probed by each geometry can vary by hundreds of Kelvin, particularly across hemispheres (from dayside 
to nightside) or even across each limb (from the colder, west terminator, to the hotter, eastern terminator; see Fig.\ref{uvt_plots}).  
Furthermore, transit observations probe shallower pressures than eclipse observations, and phase curves and eclipse maps 
can probe a range of pressures in a single dataset.  

\begin{figure*}
   \centering
   \includegraphics[trim = 0.3in 0.1in 0.3in 0.0in, clip, width=0.95\textwidth]{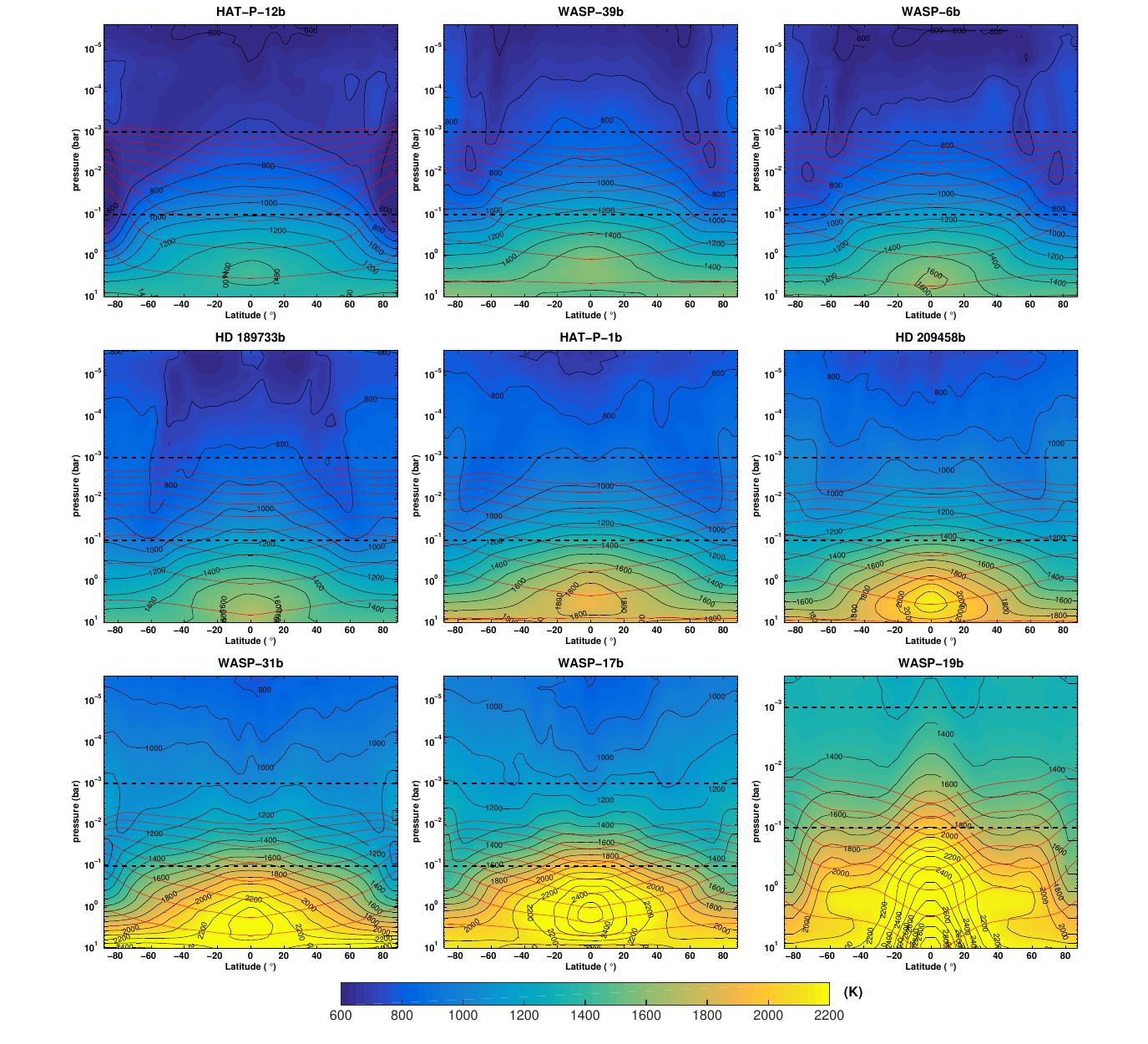}
   \caption{Limb-averaged temperature as a function of latitude and pressure for each individual LP planet. Panels are organized by increasing planetary equilibrium temperature, ${T_{eq}}$, and are plotted on the same colorscale.  These limb-averaged temperatures are calculated by weighting each grid point that is located along the limb by their grid cell length.  Black dashed lines indicate the pressures typically probed in transmission, while red
   solid lines are isentopes (lines of constant entropy).}\label{temp_lat_maps}
\end{figure*}

These considerations are important when inferring molecular abundances and cloud or haze properties for an individual planet 
or a range of planets.  For example, the water abundance retrieved from a transmission spectrum could be 
largely different in an emission spectrum, as they probe different longitudes, temperatures and pressures.  Cloud properties are 
also expected to vary across the planet with temperature, and contribute varying amounts to each spectrum.  
Therefore, the three-dimensional temperature structure must be considered when interpreting an ensemble of hot Jupiter observations, 
particularly when inferring molecular abundances and cloud properties.  Here we use models from our nine-planet study to 
illustrate the spatial temperature variations that should be considered in emission and transmission observations.

%%%%%%%%%%%%%%%%%%%%%%%%%%%%%%%%%%%%%%%%%%%%%%%%%%%%%%%%%%%%%%

\subsubsection{Emission observations: temperature variations with longitude}\label{subsection:lontemp}

When observing the thermal emission from a transiting exoplanet, the contribution of flux from the equator is 
greater than the contribution from the poles, so we can plot the temperature weighted by the cosine of the planet 
latitude, $\phi$ for insights to those observations.  This is shown in Fig. \ref{temp_lon_maps} for each LP 
planet as a function of longitude and pressure, all plotted on the same colorscale.   At a given pressure, we again 
recover the trend of increasing day-night temperature contrast with increasing $T_{eq}$, as seen in Fig. \ref{uvt_plots}.  
Furthermore, for cooler hot Jupiters HAT-P-12b, WASP-39b and WASP-6b, the hotspot is shifted approximately $\sim$40-50 
degrees eastward from the substellar point at the range of pressures probed in emission ($\sim$1 bar to 10 mbar; black dashed 
horizontal lines), while the hotter planets HD 189733b, HAT-P-1b, HD 209458b, WASP-31b, WASP-17b, and WASP-19b 
have hotspot shifts of $\sim$20-30 degrees.  

We can use these plots as a qualitative basis for predicting the behavior of their IR phase curves.  First, we can
infer that amplitudes of IR flux will likely be largest and peak IR fluxes will occur closest to 
secondary eclipse for the hottest planets (e.g., WASP-19b, WASP-17b), while cooler planets will likely exhibit smaller 
amplitudes and IR fluxes that peak well before secondary eclipse. While we do not include the theoretical phase curves 
for each planet in this paper, we do include a table of measured phase offsets and flux amplitudes derived from our models at the Spitzer IRAC 
3.6 and 4.5 $\mu$m bandpasses and the HST/WFC3 G141 bandpass (1.12-1.65 $\mu$m) (Table \ref{LP_lc_data}).  Phase offsets are
measured in units of phase relative to secondary eclipse, where secondary eclipse occurs at a phase of 0.5. 
At each bandpass, the timing of peak IR flux offset generally becomes shorter and the phase amplitude generally
becomes larger with increasing $T_{eq}$.  This is likely because radiative timescales for these hotter planets are much shorter 
than the timescales for Kelvin and Rossby waves to propagate, which likely suppress the formation of zonal jets 
\citep[e.g.,][]{showman+2013,komacek+showman_2016}.  However, chemistry also likely plays a role \citep{zellem+2014}.
We also note that our model phase curves have previously been compared to Spitzer phase curves of HD 209458b and HD 189733b 
\citep{knutson+2012,zellem+2014}, and future work will also present comparisons to Spitzer 3.6 and 4.5 $\mu$m 
phase curves of WASP-19b \citep{wong+2015b}.  We also compare theoretical dayside emission spectra to 
Spitzer observations in Section \ref{section:obsconstraints}. 

\begin{deluxetable*}{lcccccc}
\tabletypesize{\footnotesize}
\tablecaption{Maximum flux peak offsets and flux amplitudes measured from our theoretical phase curves for each LP planet, in the Spitzer 3.6 and 4.5 $\mu$m bandpasses and the HST/WFC3 bandpass (1.12-1.65 $\mu$m).}
\tablewidth{0pt}
\tablehead{
\colhead{}& \multicolumn{2}{c}{Spitzer 3.6 $\mu$m} & \multicolumn{2}{c}{Spitzer 4.5 $\mu$m} & \multicolumn{2}{c}{HST/WFC3 (1.12-1.65 $\mu$m)} \\
\colhead{Planet} & \colhead{Max offset (phase)} & \colhead{Amplitude (ppm)} & \colhead{Max offset (phase)} & \colhead{Amplitude (ppm)} & \colhead{Max offset (phase)} & \colhead{Amplitude (ppm)}
}
\startdata
HAT-P-12b & -0.193 & 138 & -0.220 & 130 & -0.251 & 3.45 \\
WASP-39b & -0.190 & 433 & -0.184 & 332 & -0.218 & 14\\
WASP-6b & -0.164 & 617 & -0.162 & 497 & -0.185 & 25\\
HD 189733b & -0.142& 950 & -0.136& 821 & -0.181 & 50\\
HAT-P-1b& -0.186 & 346 & -0.164 & 307 & -0.219 & 16 \\
WASP-31b & -0.169 & 437 & -0.144 & 620 & -0.190 & 57 \\
WASP-17b & -0.137 & 544 & -0.115 & 791 & -0.157 & 110\\
WASP-19b & -0.054 & 3391 & -0.049 & 4268 & -0.064 & 1387 
\enddata
\label{LP_lc_data}
\end{deluxetable*}

%%%%%%%%%%%%%%%%%%%%%%%%%%%%%%%%%%%%%%%%%%%%%%%%%%%%%%%%%%%%%%%%%%%%%%%%%%%%%

\subsubsection{Transmission observations: temperature variations with latitude}\label{subsection:lattemp}

Unlike emission, where the observed flux is weighted towards the equator, all latitudes have equal weighting on transit observations. 
Because we are observing only the limb, we can instead investigate the variations in limb-averaged temperature with latitude.  
These maps are shown for each LP planet in Figure \ref{temp_lat_maps} as a function of pressure and latitude.  The limb temperature
is calculated by weighting each grid point located along the limb by their corresponding grid cell length.
For each planet, the hottest regions are confined to the lowest latitudes (less than approximately $\pm30^{\circ}$), and maximum temperatures 
increase with increasing ${T_{eq}}$.  If we focus on the pressure range probed in transmission ($\sim$100-1 mbar; black dashed 
horizontal lines), we see that at a given pressure, temperature variations from the equator to pole can exceed hundreds of K for 
even the coolest hot Jupiters. On WASP-6b, for example, the temperature at 10 mbar varies from approximately 1200 K at the 
equator to approximately 800 K at the poles, a difference of $\sim$400 K.  

The slope of the temperature contours for HAT-P-12b are large, consistent with the large equator-to-pole temperature 
differences seen in Fig. \ref{uvt_plots}. Coincident with these sloping contours are sloping isentropes (red contours), 
which indicate constant entropy surfaces.  The isentrope slopes of HAT-P-12b become particularly large at high latitudes at pressures of 
1 bar to 10 mbar, and suggest that the atmosphere is dynamically unstable \citep{showman+2015}.  
Indeed, this instability would further explain the high-latitude jets seen in HAT-P-12b's zonal-mean zonal wind profile (Fig. \ref{uz_plots}), 
as it would lead to development of baroclinic eddies at mid- to high-latitudes, which transport heat meridionally and force and 
maintain jets at those latitudes.  The variably sloping isentropes on WASP-19b suggest baroclinic eddies also play a role in the 
formation of their high-latitude jets, particularly when coupled with its small deformation radius.  Among the rest of the nine-planet 
sample, the isentrope slopes are comparatively flat, suggesting their atmospheres are comparatively stable.  This is especially true 
for those hot Jupiters that are highly irradiated and slowly rotating, such as WASP-17b and HD 209458b \citep[see Table 
\ref{LP_params} and][]{showman+2015}.

\begin{figure*}
   \centering
   \includegraphics[trim = 0.0in 0.2in 0.0in 0.2in, clip, width=1.0\textwidth]{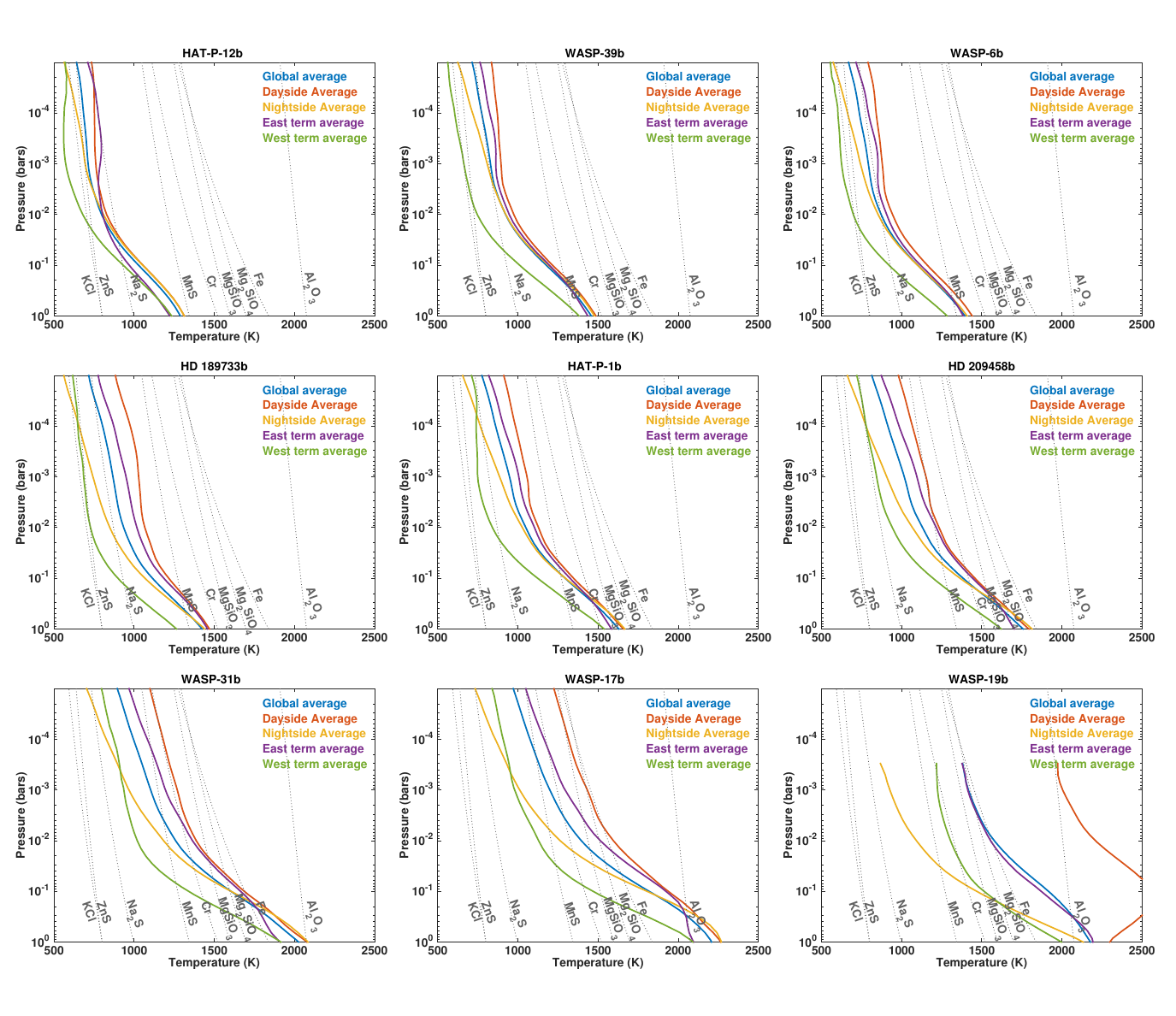}
   \caption{Global- dayside-, nightside-, east terminator- and west terminator-averaged temperature as a function of pressure for each individual LP target.  Panels are organized by increasing planetary equilibrium temperature, ${T_{eq}}$.  Condensation curves are also plotted in grey dotted lines.}\label{1Dprofs}
\end{figure*}

\subsubsection{Comparing 1D P-T profile averages}
We can summarize the three-dimensional variations in temperature structure for each planet using one-dimensional averages, 
particularly over each hemisphere and limb.  Fig. \ref{1Dprofs} shows the dayside-, nightside-, global-, east limb- 
and west limb-averaged temperature as a function of pressure for each individual planet.  Comparing only the dayside- 
and nightside-averaged profiles (red and yellow profiles, respectively) we see that consistent with our wind and 
temperature profiles in Figure \ref{uvt_plots}, the difference between dayside- and nightside-averaged temperature
 is largest for the hottest planet, WASP-19b, with a difference of $\sim$1000 K at 1 mbar, and smallest for the coolest planet, HAT-P-12b,
 where the temperature difference is only $\sim$50 K.  
 
 When we decompose the limb-averaged profile 
mapped in Figure \ref{temp_lat_maps} to their east and west limb averages (purple and green profiles, respectively), 
we see that for all nine planets at mbar pressures, the eastern limb is significantly hotter than the west limb.  Interestingly,
 with increasing $T_{eq}$, differences in west and east limb temperatures
at photospheric pressures increase then decrease, with WASP-17b, not WASP-19b, having the largest temperature 
differences ($\sim$300-400 K).  This
is likely because the increase in overall temperature is mediated by the reduced hotspot offset and increased day-night flow, which
begins to homogenize the east- and west-limb temperatures for WASP-19b.

This range of temperature differences between profiles and 
planets, can imply a diversity in molecular abundances, as well as cloud compositions and cloud base pressures, which will discuss
in the following sections.  We also note that for all but the hottest planets, the west limb reaches the coolest temperatures, 
and are actually significantly cooler than even the nightside averages.

\begin{figure*}
\centering
   \includegraphics[trim = 0.0in 0.1in 0.0in 0.1in, clip, width=1.0\textwidth]{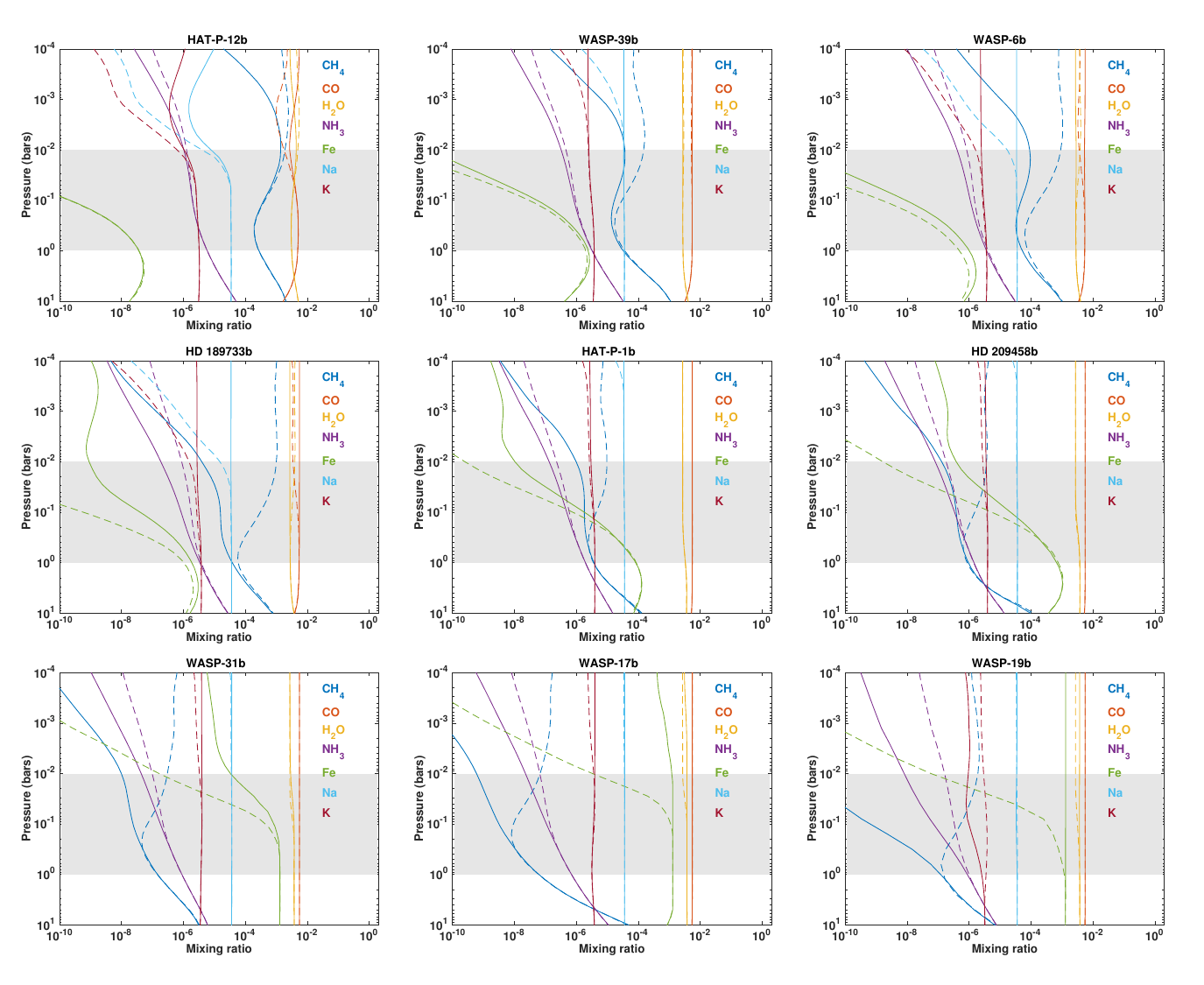}
   \caption{Chemical equilibrium abundances across the dayside (solid profiles) and nightside (dashed profiles) for each LP planet.  ${\rm CH_4}$ and CO abundances are indicated by the thicker orange and blue lines, respectively.  Grey boxes indicate the pressures typically probed in emission.}\label{abunds_daynight}
\end{figure*}

\begin{figure*}
\centering
  \includegraphics[trim = 0.0in 0.1in 0.0in 0.1in, clip, width=1.0\textwidth]{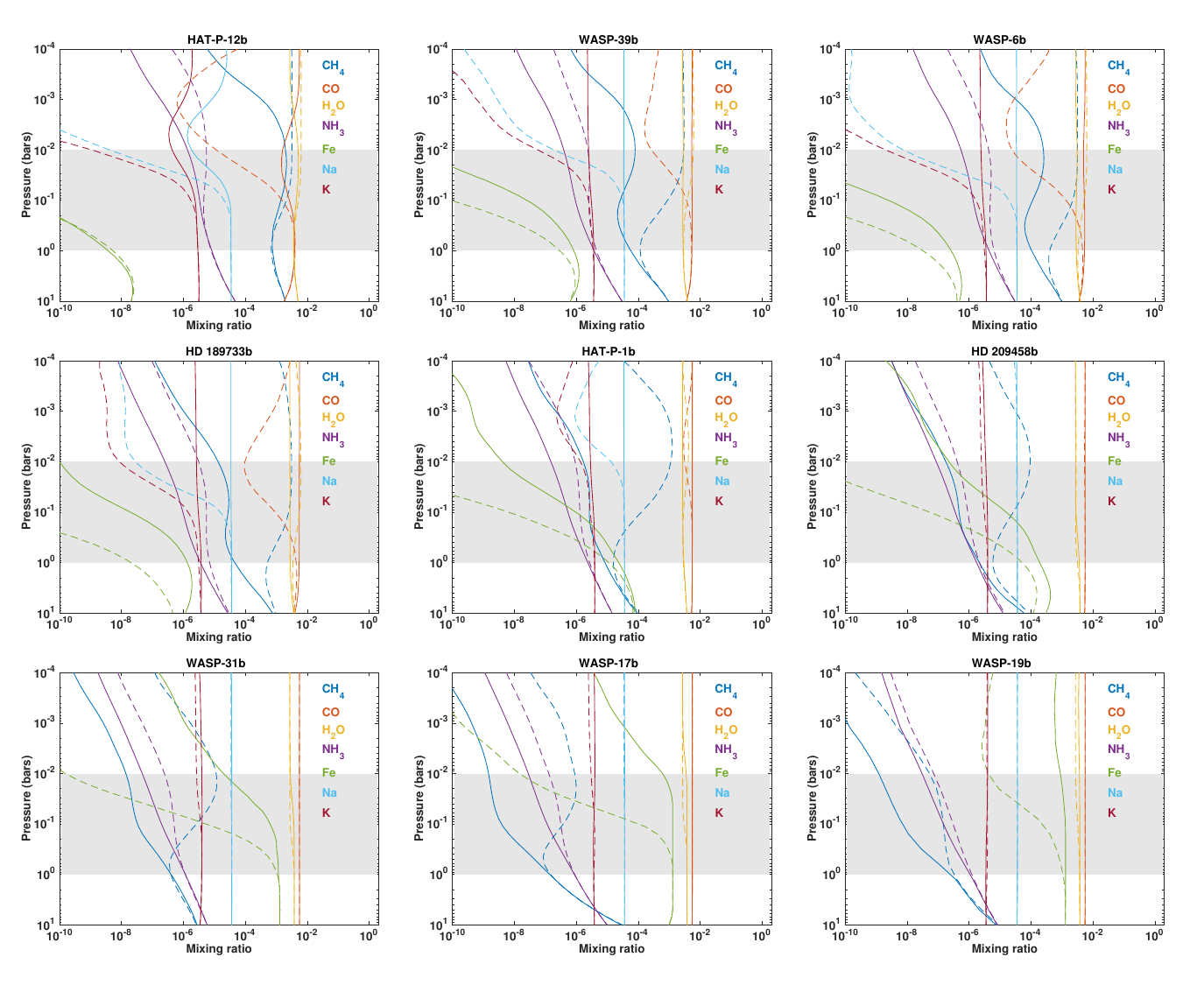}
  \caption{Chemical equilibrium abundances across the eastern (solid profiles) and western (dashed profiles) limbs for each LP planet.  ${\rm CH_4}$ and CO abundances are indicated by the thicker orange and blue lines, respectively. Grey boxes indicate the pressures typically probed in transmission.}\label{abunds_eastwest}
\end{figure*}

\begin{figure*}
   \centering
   \includegraphics[trim = 0.05in 0.1in 0.05in 0.0in, clip, width=1.0\textwidth]{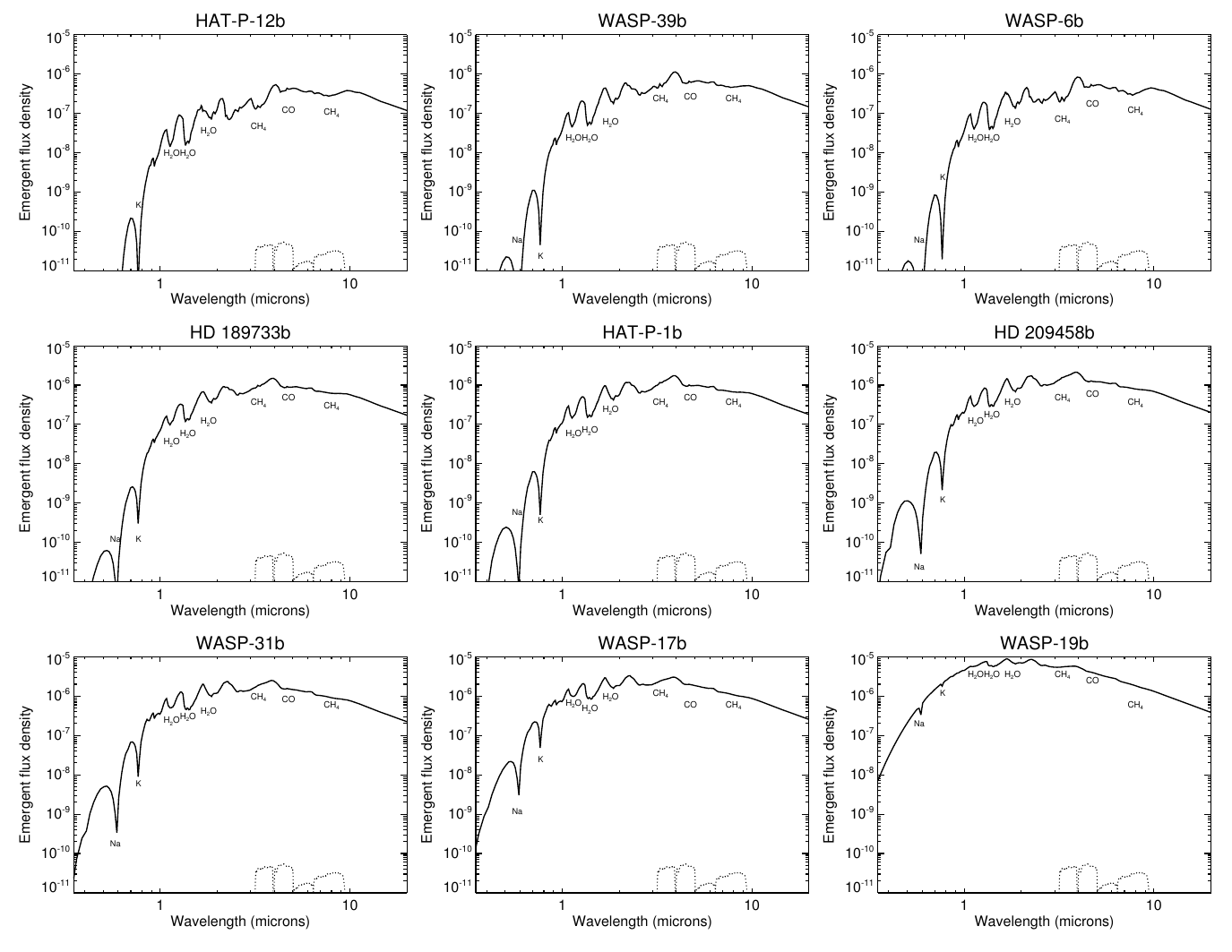}
    \caption{Dayside emergent flux density (ergs ${\rm s^{-1}~cm^{-2}~Hz^{-1}}$) vs. wavelength for each LP target.  Normalized transmission functions for Spitzer 3.6, 4.5, 5.8 and 8.0 $\mu$m channels
   are shown in black dotted lines.}\label{ermflux}
\end{figure*}

\begin{figure*}
\centering
   \includegraphics[trim = 0.0in 0.1in 0.0in 0.0in, clip, width=1.0\textwidth]{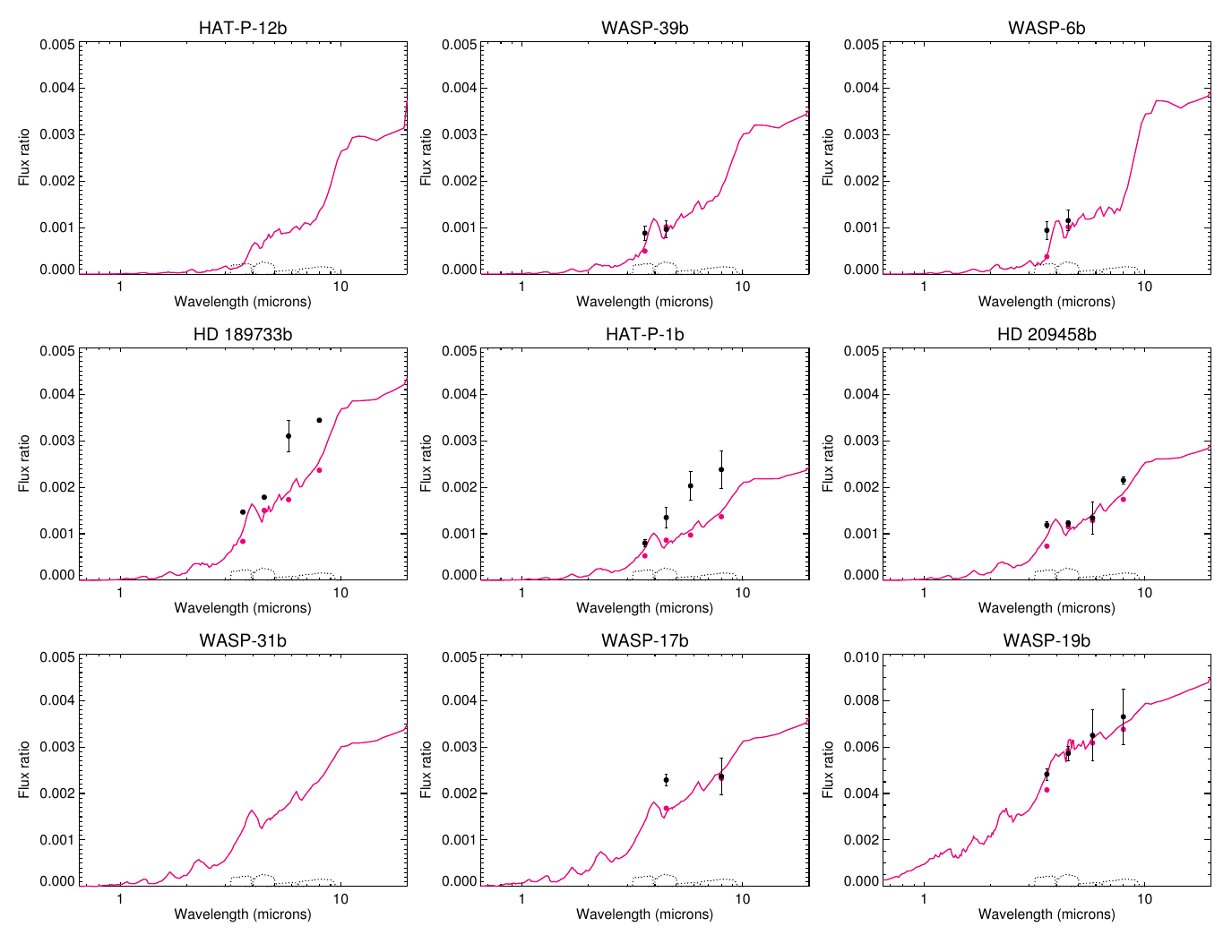}
   \caption{Planet-to-star flux ratio (pink profiles) vs. wavelength for each LP planet.  Pink dots show the flux ratio integrated over each Spitzer bandpass, whose
   normalized transmission functions are shown as black dotted lines. Overplotted in black dots with error bars are published Spitzer secondary eclipse depths for WASP-39b, WASP-6b, HD 1897433b, HAT-P-1b, HD 209458b, WASP-17b and WASP-19b.  See text for corresponding references to each dataset. }\label{emspec}
\end{figure*}

%%%%%%%%%%%%%%%%%%%%%%%%%%%%%%%%%%%%%%%%%%%%%%%%%%%%%%%%%%%%%%%%%%%%%%%%%%%%%
\subsection{Chemical implications}\label{section:chemistry}
Taken together, the spatial variations in temperature from dayside to nightside, from equator to pole, 
and also between the east and west terminators, can imply large differences in major molecular abundances.  
In this section we calculate the mass mixing ratios for seven major species ($\rm{CH_4, 
CO, H_2O, NH_3}$, Fe, Na and K) by interpolating the dayside-, nightside-, 
east- and west-terminator-averaged 1D $p$-T profiles from Figure \ref{1Dprofs} 
onto a chemical equilibrium abundance P-T grid.  While we note that the these 
highly-irradiated atmospheres may not obey local chemical equilibrium (see below), 
here we assume a pristine environment to explore these abundances.

Figures \ref{abunds_daynight} and \ref{abunds_eastwest} plot the mass mixing ratios as a function of pressure.
In Figure \ref{abunds_daynight}, solid (dashed) profiles signify the dayside (nightside) abundances, 
while the eastern (western) limb abundances are denoted by the solid (dashed) lines in Figure \ref{abunds_eastwest}.  
Rather than focusing on the abundances of each individual planet, we identify trends in abundance variations for 
each of the species (or groups of species) in Sections \ref{subsubsection:h2o}-\ref{subsubsection:co}.

\subsubsection{$H_2O$ abundance} \label{subsubsection:h2o}
$\rm{H_2O}$ is one of the more abundant species in each planet's atmosphere (yellow profiles). Its mixing ratio
remains largely unchanged with pressure for each planet in our sample; this is true for the dayside 
and nightside profiles (Fig.\ref{abunds_daynight}, yellow solid and dashed profiles, respectively), as well as the 
east and west limb profiles (Fig. \ref{abunds_eastwest}, yellow solid and dashed profiles, respectively).

\subsubsection{Fe abundance} 
Considering the dayside and nightside profiles in Figure \ref{abunds_daynight}, we would expect that Fe (green profiles)
would not be very abundant in the atmospheres of HAT-P-12b, WASP-39b, WASP-6b and HD 189733b, with approximately equal 
mixing ratios on the dayside and nightside.  As the we move to hotter planets in our sample, where 
Fe is expected to condense on the nightside, Fe abundances on the dayside and nightside can differ by orders 
of magnitude; this is true for WASP-17b, whose dayside Fe mass mixing ratio is $\sim10^{-3}$ at all pressures and 
whose nightside abundance falls from $\sim10^{-4}$ at 100 mbar to $10^{-10}$ at $\sim$1 mbar.  Comparing the 
east and west limb abundances (Fig. \ref{abunds_daynight}), Fe is more abundant on the eastern limb than the 
western limb by at least an order of magnitude at pressures probed in transmission (grey box) for all planets 
except HAT-P-12b, whose Fe abundance is equally small on both limbs.  

\subsubsection{$NH_3$ abundance} 
Comparing the dayside and nightside mass mixing ratios, ammonia abundance does not differ significantly for most planets 
in our sample (purple profiles).  The exception is WASP-19b, where ammonia is less abundant on the dayside than the nightside by 
1-2 orders of magnitude at pressures probed in emission.  Differences in ammonia abundance are more 
noticeable when comparing the eastern and western limbs; between 1-100 mbar the variations are about 
an order of magnitude.  However, those variations are actually smallest for WASP-19b.  

\subsubsection{Na and K abundance} 
Comparing the dayside and nightside mass mixing ratios (Fig. \ref{abunds_daynight}), the abundances of Na and K (light blue and 
crimson profiles, respectively) are approximately constant with pressure for planets with ${T_{eq}\agt1300}$ K: HAT-P-1b, HD 209458b, 
WASP-31b, WASP-17b and WASP-19b, and do not 
differ significantly between the dayside and nightside.  However, for the four cooler planets with ${T_{eq}\alt1200}$ K (HAT-P-12b, WASP-39b, 
WASP-6b, and HD 189733b), Na and K nightside abundances deviate from their dayside abundances at pressures 
less than $\sim10$ mbar, where they are depleted by the formation of $\rm{Na_2S}$ and KCl clouds.  In the case of WASP-19b, K is actually
more abundant on the nightside; this is because dayside temperatures are high enough for an appreciable fraction of K to ionize.

Comparing the eastern and western terminator abundances (Figure \ref{abunds_eastwest}), we see the same 
general trend, with hotter planets largely abundant in Na and K and minimal deviations between east and west 
limbs.  However, HAT-P-1b is among the cooler planets that show significant differences in east/west limb 
abundances; the western limb Na and K abundances differ from the eastern limb by approximately two orders of 
magnitude at pressures of $\sim$1 mbar.  Planets cooler than HAT-P-1b show deviations of at least four orders 
of magnitude between east and west limb abundances at pressures less than $\sim$10 mbar.  These cooler 
planets, then, would be expected to have reduced Na and K transmission signals on the western terminator
as compared to the comparatively hotter targets, due to the formation of clouds.  
We further discuss the presence of clouds and the potential to probe these abundance
variations in transit in Section \ref{section:discussion}.

\subsubsection{$CO$ and $CH_4$ abundance}\label{subsubsection:co}
This nine-planet sample is especially interesting for comparing CO and $\rm{CH_4}$ abundances (orange and 
blue profiles, respectively), as the intraconversion between these two species occurs over this temperature range, 
similar to the L to T transition for brown dwarfs.  
If we first consider only the dayside profiles in Figure \ref{abunds_daynight}, we see that the dayside CO 
varies little with pressure for each planet in our sample.  Furthermore, for all planets 
except HAT-P-12b, $\rm{CH_4}$ is less abundant than CO on the dayside by at least 2-4 orders of magnitude.  
Comparing only the nightside profiles, we see that nightside $\rm{CH_4}$ abundance increases with decreasing 
${T_{eq}}$, consistent with theoretical expectations.  These nightside abundances, however, are still orders 
of magnitude smaller than nightside CO abundances for WASP-39b, HAT-P-1b, HD 209458b, WASP-31b, 
WASP-17b, and WASP-19b. 

The dayside and nightside $\rm{CH_4}$ abundances are directly related to the trends in day-night temperature 
variation with increasing ${T_{eq}}$ (Figs. \ref{uvt_plots} and \ref{1Dprofs}).  WASP-19b, with its large day-night 
temperature contrast, exhibits the largest differences in day-night $\rm{CH_4}$ abundances, with $\rm{CH_4}$ 
more than four orders of magnitude more abundant on the nightside than the dayside (bottom right panel).  
Conversely, for planets with small day-night temperature contrasts (e.g., WASP-39b, HAT-P-12b), the differences 
in day-night $\rm{CH_4}$ abundance are much smaller.  For HAT-P-12b, WASP-6b and HD 189733b, $\rm{CH_4}$
becomes nearly equal in abundance to CO on the nightside, particularly at pressures near 
1 mbar.  For HAT-P-12b, methane is actually more abundant than CO on the nightside at pressures less than 1 mbar.  

The differences in $\rm{CO/CH_4}$ abundance are more dramatic when considering the east and west terminators 
(Fig. \ref{abunds_eastwest}).  Differences between east- and west-limb $\rm{CH_4}$ abundances are largest for 
those with the largest differences in limb temperatures (see Fig \ref{1Dprofs}). For the hottest planets in our sample 
(HAT-P-1b, HD 209458b, WASP-31b, WASP-17b, and WASP-19b), CO is the dominant carbon-bearing molecule on 
both the western and eastern limbs, particularly at the mbar pressures probed in transmission.  However, for cooler 
planets (WASP-6b, WASP-39b, HD 189733b) while CO is more abundant on the eastern limb by at least an order of 
magnitude, methane is more abundant than CO on the western limb by at least 1-2 orders of magnitude.  In the case 
of HAT-P-12b, the coolest planet in our sample, methane is more abundant than CO on the western limb by 3-4 orders
 of magnitude, and methane and CO have nearly equal abundances on the eastern limb.  These large deviations in 
 molecular abundances occur at pressures below $\sim$100 mbar. 
 
However, it is likely that for all but the hottest planets in our sample (WASP-17b and WASP-19b), 
 CO and $\rm{CH_4}$ abundances may be out of equilibrium on the dayside, nightside, and limbs 
 \citep{cooper+showman_2006,agundez+2012,agundez+2014}.  
\cite{cooper+showman_2006} have shown that that for temperatures $\le$2000 K, 
the $\rm{CO/CH_4}$ intraconversion timescale at these temperatures and pressures
is long ($\gg10^5$ s).  Horizontal and vertical advective timescales are comparatively short, and therefore the dynamics
 will likely force the CO and $\rm{CH_4}$ abundances to be fixed with pressure.  

Overall, the wide variations in temperature with longitude and latitude can imply significant differences in chemical 
abundances, particularly with regards to Na, K, CO and $\rm{CH_4}$ abundances.  We discuss their significance further, 
as well as their observational prospects, in Section \ref{section:discussion}.

\begin{figure}
\centering
   \includegraphics[trim = 0.0in 0.0in 0.0in 0.0in, clip, width=0.5\textwidth]{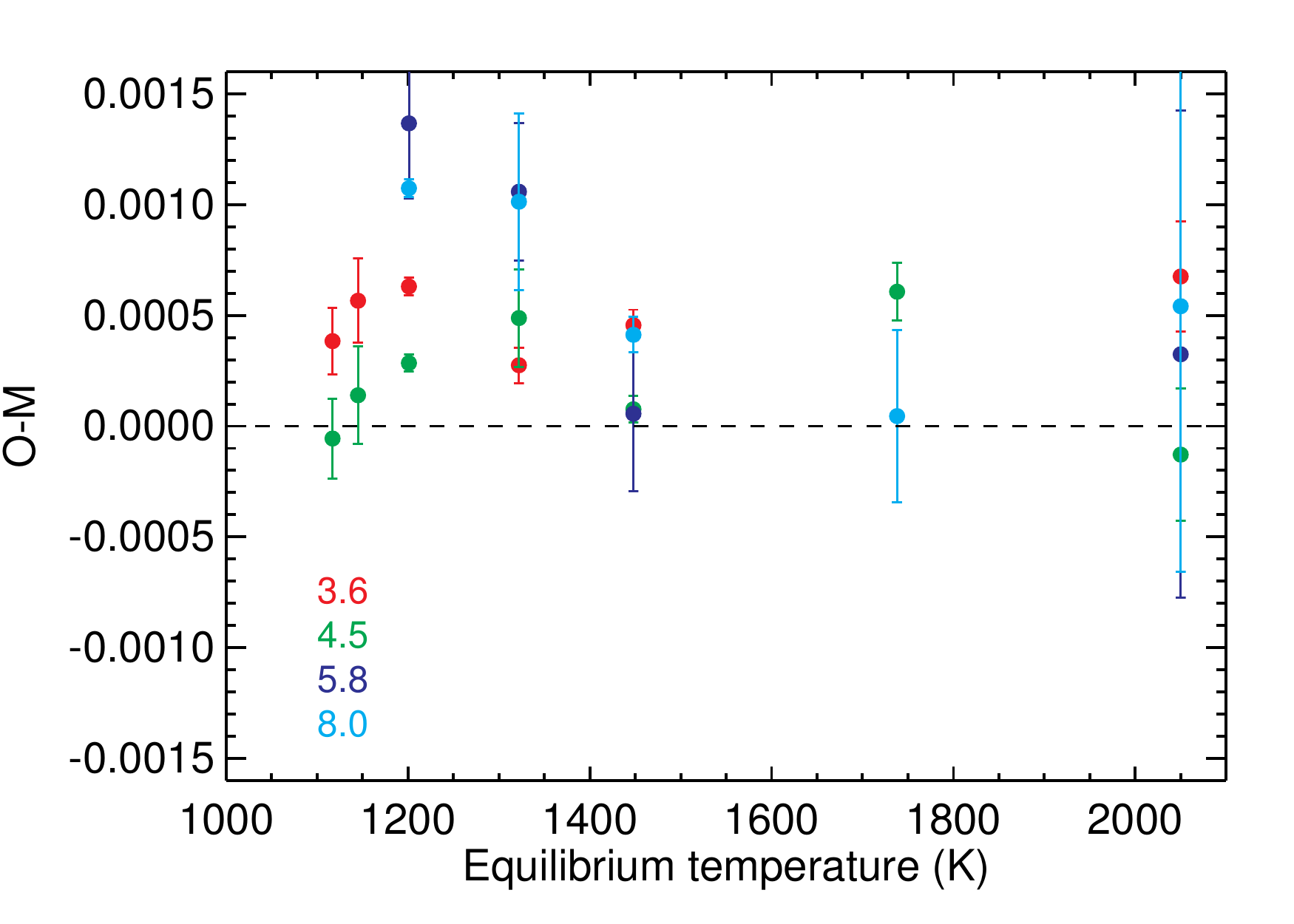}
   \caption{Difference in observed and model eclipse depths as a function of equilibrium temperature, ${T_{eq}}$.  Red, green, navy and light blue
   points, with error bars, correspond to eclipse depth differences at 3.6, 4.5, 5.8 and 8.0 $\mu$m.}\label{eclipse_trend}
\end{figure}

%%%%%%%%%%%%%%%%%%%%%%%%%%%%%%%%%%%%%%%%%%%%%%%%%%%%%%%%%%%%%%%%%%%%%%%%%%%%
\subsection{Comparison to Spitzer eclipse data}\label{section:obsconstraints}
Using our nine-planet sample, we can look for trends not only in temperature and molecular abundances, but also 
how that translates to trends in observations.  We focus primarily on observations in emission, as they are more 
directly related to each planet's thermal structure, and much less sensitive to the presence of clouds.  We show 
theoretical dayside emission spectra generated from our circulation models in Figure \ref{ermflux}.  For all planets, 
absorption features due to Na, K, $\rm{CH_4}$ and ${\rm H_2O}$ are apparent.  These spectral features 
 deepen as we probe the upper atmospheres of the cooler planets
(compare, for example, the features for WASP-6b as compared to the features in WASP-17b).  
 With its large mixing ratio, CO is expected to be abundant across the sample, but its spectral features are less prominent due to its smaller absorption cross section.  

We can then compare our models to available secondary eclipse data by calculating the ratio of the planetary flux to the
stellar flux for each system (Figure \ref{emspec}).  The model spectra are plotted in pink profiles, with the binned fluxes for each Spitzer
channel in pink dots.  Transmission curves for each Spitzer channel are shown as black dotted curves.  
Overplotted in black dots with error bars are published Spitzer secondary eclipse depths for WASP-39b \citep{kammer+2015}, 
WASP-6b \citep{kammer+2015}, HD 1897433b \citep{charb+2008,agol+2010,knutson+2012}, 
HAT-P-1b \citep{todorov+2010}, HD 209458b \citep{diamondlowe+2014}, 
WASP-17b \citep{anderson+2011} and WASP-19b \citep{anderson+2013}.  

Overall, each 1$\times$ solar model spectrum is in good general agreement with each Spitzer dataset.  
This is noteworthy for a number of reasons.  First, while comparative circulation studies to date have
only explored synthetic observations for one or two planets models over a wide
phase space \citep[e.g.,][]{showman+2009,kataria+2015,showman+2015}, our study is the 
first to present synthetic observations from nine individual planet models, with their individual system parameters.  
Second, we generate theoretical observations from those models without any additional tuning (e.g., by varying drag,
C/O ratio, chemistry, clouds, etc.).  The fact that these purely three-dimensional outputs are in generally good agreement over
over such a wide phase space is notable.  

Still, there are discrepancies between the models and the observations, mostly notably the 5.8 and 8 micron 
points for HD 189733b, the 4.5 and 8 micron points for HAT-P-1b, and the 4.5 micron eclipse depth for WASP-17b.  
In Figure \ref{eclipse_trend}, we quantify the agreement by differencing the observed and model eclipse depths 
for each Spitzer observation (black and pink points, respectively), and plot them as a function of ${T_{eq}}$.  
The variation of the 3.6, 4.5, 5.8 and 8.0 $\mu$m
points (red, green, navy, and light blue points, respectively) with ${T_{eq}}$ does not appear to follow any clear trend,
suggesting that our model disagreement does not correlate with equilibrium temperature.

It is interesting to note that the points with the least agreement 
are those with the oldest reductions and some of the largest uncertainties (e.g., HAT-P-1b).
We look forward to a re-analysis of these data to determine if revisions in eclipse depths are warranted, 
as has been the case for HD 189733b and HD 209458b \citep{knutson+2012,diamondlowe+2014,zellem+2014,evans+2015}. Furthermore,
our models seem to agree best with the 4.5 $\mu$m points, where we are probing the abundance of CO, which is not 
expected to vary widely with $T_{eq}$ (Figure \ref{abunds_daynight}).

Nevertheless, Figure \ref{eclipse_trend} shows that our circulation models generally underpredict the dayside flux, 
although only on the $\sim$100 ppm level (with uncertainties).  This underprediction could be due to differences 
in atmospheric metallicity.  For example, \cite{knutson+2012} show that the Spitzer dayside photometry of 
HD 189733b is well-matched by a model with an atmospheric composition of 5$\times$ solar.  Furthermore, 
preliminary analyses of Spitzer phase curves of WASP-19b at 3.6 and 4.5 $\mu$m show that an enhanced
metallicity atmosphere (also 5$\times$ solar) is in better agreement with the amplitude of IR flux and timing of peak IR flux \citep{wong+2015b}.
Magnetic drag or high C/O ratios could also enhance dayside fluxes and provide better agreement.

%%%%%%%%%%%%%%%%%%%%%%%%%%%%%%%%%%%%%%%%%%%%%%%%%%%%%%%%%%%%%%
\section{Discussion}\label{section:discussion}

Using the circulation models for our nine-planet sample, we have shown that over a wide range of 
planet radii, gravities, orbital periods, and equilibrium temperatures, equatorial superrotation 
continues to be a robust dynamical feature for hot Jupiters, and results in the eastward displacement 
of the dayside hotspot at photospheric pressures. Our results from Table \ref{LP_lc_data} 
suggest that with increasing $T_{eq}$, hot Jupiters 
exhibit larger phase amplitudes and peak IR flux offsets that occur closer to secondary eclipse. 
While phase curve observations have only been collected for a few
hot Jupiters \citep[see][and references therein]{wong+2015a}, we look forward to future 
phase curve observations with Spitzer, HST, K2, and JWST to further probe $T_{eq}$ space.

Although the focus of this study is to compare the planet sample in its entirety, we note that our models of 
HD 189733b and HD 209458b are qualitatively similar to circulation models by ours and other groups 
\citep[e.g.,][]{showman+2009,dobbsdixon+2010,heng+2011,dobbsdixon+agol_2013,kataria+2013,liu13,mayne+2013,
parmentier+2013,perna+2010,rauscher+menou_2013,rogers+komacek_2014,rogers+showman_2014}.  
Each model exhibits equatorial superrotation with peak speeds on the order of 1 ${\rm km~s^{-1}}$, and has 
an eastward-shifted ($\sim20-40^{\circ}$) hotspot at photospheric pressures.  The differences between models
 could arise in part from differences in the dynamical core: 
\cite{dobbsdixon+2010}, \cite{dobbsdixon+agol_2013} and \cite{mayne+2013}, for example, employ the fully 
compressible, 3D Navier-Stokes equations, while \cite{rogers+komacek_2014} and \cite{rogers+showman_2014} 
employ a 3D magnetohydrodnamics (MHD) model in the anelastic approximation.  Differing radiative transfer 
schemes also lead to differences in heating and cooling and therefore day-night temperature contrasts and 
equatorial jet speeds; \cite{dobbsdixon+2010} and \cite{rauscher+menou_2013} utilize a dual-band radiative 
transfer model in modeling HD 189733b and HD 209458b, while \cite{heng+2011}, \cite{mayne+2013}, 
\cite{rogers+komacek_2014} and \cite{rogers+showman_2014} use a Newtonian cooling scheme to model 
HD 209458b.  Lastly, some models include the effects of either magnetic drag and/or ohmic dissipation, 
which serve to reduce day-night temperature contrasts and equatorial jet speeds 
\citep{perna+2010,rauscher+menou_2013,rogers+komacek_2014,rogers+showman_2014}.

Our results in Section \ref{section:chemistry} demonstrate that one must be careful in extrapolating 
atmospheric chemical abundances based on a single observational technique.  We cannot assume, 
for example, that the constrained abundances from an emission spectrum will be the same abundances 
as a transmission spectrum, even for cloud-free atmospheres, as each set of observations probes 
difference longitudes, pressures, and chemical regimes.  Figures \ref{abunds_daynight} and 
\ref{abunds_eastwest} show that these abundances can vary by many orders of magnitude, especially 
when comparing the abundances on the dayside and terminators.  As the fidelity of exoplanet 
observations improve, one must be more careful to consider these factors particularly when combining 
data sets to constrain global thermal and chemical properties of a given planet.  

High-dispersion ($R\equiv\lambda/\Delta\lambda\sim $100,000) spectrographs have the potential to probe
the spatial chemical variations shown in our models.  Already, these instruments have demonstrated the 
ability to detect and resolve Na, K, CO, ${\rm CH_4}$ and ${\rm H_2O}$ spectral lines for transiting hot Jupiters
in transmission \citep{redfield+2008,snellen+2008,snellen+2010,
wyttenbach+2015,louden+wheatley_2015,brogi+2015}, and for transiting {\it and} non-transiting hot Jupiters in emission 
\citep{rodler+2012,birkby+2013,dekok+2013,brogi+2012,brogi+2013,brogi+2014,schwarz+2015}.  
Future high-resolution spectrographs such as ESPRESSO or CRIRES+ 
on the Very Large Telescope (VLT), or HiReS and METIS on the Extremely Extra Large Telescope (E-ELT), 
would be well-suited to this task.  The JWST/MIRI and NIRSpec instruments could potentially probe and 
resolve these spectral variations as well. 

Such measurements could focus solely on Na and K, as has already been done for HD 189733b and HD 209458b
\citep[e.g.,][]{redfield+2008,snellen+2008,wyttenbach+2015,louden+wheatley_2015}.  
We would expect that for the cooler planets in our sample, Na and K would be less abundant on the western limb and 
nightside, which would translate to a smaller transmission signal.  However, it should be noted that 
other Na- and K-bearing species become increasingly abundant at lower
temperatures, such as NaCl, NaOH and KOH \citep{lodders1999}. While monatomic Na will 
tend to dominate over most conditions, monatomic K may be depleted by the formation of these species.  
Furthermore, it is possible that cold traps \citep[e.g.,][]{parmentier+2013} 
could also suppress variations in Na, K, and Fe abundance.  

HAT-P-12b could be an ideal testbed for probing differences in CO/$\rm{CH_4}$ abundances, 
as the planet could exhibit appreciable differences in these abundances between the east and 
west limbs, and could have equal abundances on the dayside, 
while also having a favorable brightness contrast compared to smaller, dimmer 
sub-Neptunes and super Earths.  However, as noted earlier, it is possible that these 
carbon-bearing species would be out of equilibrium, which would freeze their abundances 
across the dayside, nightside and limbs.  In this case, the differences in CO and ${\rm CH_4}$ 
abundance would be largely reduced \citep{agundez+2012,agundez+2014,cooper+showman_2006}.
  We look forward to future observations to test these predictions.

The interpretation of these future datasets could also be complicated by the formation of clouds, as evidenced by our sample of transmission spectra 
\citep{sing+2016}.  \cite{line+parmentier_2015} have shown that partially cloudy limbs can appear as a residual in transit
ingress and egress with a residual of $\sim$100 ppm.
While we do not compare our cloud-free models directly to transmission spectra to interpret their cloud properties, we
can use our model results to explore their potential three-dimensional structure.  If we first consider the limb-averaged temperature 
maps in Figure \ref{temp_lat_maps}, it is likely that a particular cloud species could form high in the atmosphere at the 
equator and mid-latitudes, while forming at deeper pressures at high latitudes. These clouds will contribute equally to a 
transmission spectrum.  

We can also use Figure \ref{1Dprofs} to further comment on cloud properties, as each panel
 includes condensation curves for chemical species expected to condense in the 
atmospheres of our planet sample (grey dotted profiles).  Intersections between each condensation curve and $p$-T profile
indicate potential cloud bases for that particular species.  While temperature differences between 
each $p$-T profile can be small ($\sim$50-75 K), these small changes can still imply large differences in cloud properties.
For example, the hot Jupiters HAT-P-12b, WASP-6b and HD 189733b have temperatures that could 
allow for the formation of ZnS, KCl and ${\rm Na_2S}$ clouds at observable pressures, but only on the western limb and nightside 
(green and yellow profiles).  This is notable, as transmission spectra for all three targets suggest their atmospheres are cloudy.  
Western limb clouds are similar to the scenario proposed for hot Jupiter Kepler-7b, which has been postulated
to have clouds westward of its substellar point based on Spitzer and Kepler phase curves \citep{demory+2013}.
Also, nightside condensation has long been hypothesized for hot Jupiters and brown dwarfs, particularly as a 
means for removing Na via ${\rm Na_2S}$ condensation \citep[][and references therein]{visscher+2006}.
Finally, all five $p$-T profiles of WASP-31b cross numerous condensation curves at observable pressures, 
including those for silicate clouds.  This suggests that its atmosphere can host many types of clouds, consistent 
with its cloudy transmission spectrum.   

It is interesting to note that HD 189733b, WASP-39b and WASP-6b appear to have very similar wind and temperature structures,
which yield similar variations in molecular abundances and emission spectra. However, within the LP transmission
spectral survey, HD 189733b and WASP-6b appear cloudy, while WASP-39b appears comparatively clear (Sing et al. 2015).  
Overall, because these models are cloud-free, more physics is warranted to further explain the trends 
seen in the transmission spectra.  Detailed cloud models would account for variations in horizontal and vertical 
mixing, which could allow for the transport of cloud particles at depth or at varying longitudes.  High-altitude photochemical hazes
are also likely to be important, as they form at temperatures less than or equal to $\sim$1000-1100 K \citep{Zahnle+2009,moses_2014}. 
We save a detailed exploration of cloud properties for a future paper (Kataria et al., in preparation).  

%%%%%%%%%%%%%%%%%%%%%%%%%%%%%%%%%%%%%%%%%%%%%%%%%%%%%%%%%%%%%%%%%%%%%%%%%%%%%

\section{Conclusions}\label{section:conclusions}
We present three-dimensional circulation models for a ``grid" of nine hot Jupiters 
that comprise a transmission spectral survey using the Hubble and Spitzer Space Telescopes, 
the largest circulation comparison study conducted for hot Jupiters to date.  We utilize these models as a
first step to interpreting their transmission spectra, which imply a large diversity in clouds and haze properties.

Across the wide range of parameters for each individual system, we show that each planet atmosphere 
exhibits equatorial superrotation, with eastward displacement of the hottest regions
from the substellar longitude.  We also show that variations in temperature, particularly from the dayside to nightside, 
and across the western and eastern limbs, can produce large variations in chemistry.  
These temperature variations could also imply large variations in cloud properties 
across those regions, though we do not investigate those aspects in detail here.  

Transmission observations have the greatest potential for probing these chemical variations, especially with the 
high-resolution spectrographs aboard the next generation of telescopic facilities (e.g., VLT, E-ELT, JWST).
Indeed, even current RV instruments such as the HARPS spectrograph have demonstrated the potential 
to probe variations in Na and K abundances \citep{wyttenbach+2015,louden+wheatley_2015}.

In comparing synthetic emission spectra for each planet with available Spitzer eclipse observations, we
find that while our solar metallicity, drag and cloud-free models agree reasonably well to observations, they do
systematically underpredict the measured dayside fluxes.  This could point to enhanced metallicity or drag, which would
 help to increase the day/night temperature contrast and therefore the dayside emergent flux.  If such scenarios are the case,
that could also imply the abundance variations discussed here would be further amplified.  In summary,
the hot Jupiters continue to be an exoplanet population rife for characterization and further understanding.  The lessons we learn about 
their chemistry and clouds are valuable for informing future observations of Neptunes, super Earths, and habitable exoplanets. 
%%%%%%%%%%%%%%%%%%%%%%%%%%%%%%%%%%%%%%%%%%%%%%%%%%%%%%%%%%%%%%%%%%%%%%%%%%%%%

\acknowledgments
We thank Caroline Morley, Nikolay Nikolov, Thomas Evans and Hannah Wakeford for useful discussions which improved the manuscript.  
We also acknowledge that part of this work was completed at the Space Telescope Science Institute (STScI) operated by AURA, Inc.
This work is based on observations with the NASA/ESA HST, 
obtained at the STScI. This work is also based in part on 
observations made with the Spitzer Space Telescope, which is operated by the Jet Propulsion Laboratory, California 
Institute of Technology under a contract with NASA. The research leading to these results has received funding from the 
European Research Council under the European UnionÕs Seventh Framework Program (FP7/2007-2013) / ERC grant 
agreement no. 336792.  Support for this work was provided by NASA through grants under the HST-GO-12473 program 
from the STScI.  This work used the DiRAC Complexity system, operated by the University of Leicester IT Services, which 
forms part of the STFC DiRAC HPC Facility (www.dirac.ac.uk). This equipment is funded by BIS National E-Infrastructure 
capital grant ST/K000373/1 and STFC DiRAC Operations grant ST/K0003259/1. DiRAC is part of the National E-Infrastructure.  
We also thank the anonymous referee for comments that greatly improved the paper.

\bibliographystyle{apj}
\bibliography{ms}

\end{document}